# "Plasmonics" in free space: observation of giant wavevectors, vortices and energy backflow in superoscillatory optical fields

**Guang Hui Yuan[1#], Edward T. F. Rogers[2,3], and Nikolay I. Zheludev[1,2*]**

[1]Centre for Disruptive Photonic Technologies, TPI, SPMS, Nanyang Technological University, Singapore 637371, Singapore
[2]Optoelectronics Research Centre and Centre for Photonic Metamaterials, University of Southampton, Highfield, Southampton, SO17 1BJ, UK
[3]Institute for Life Sciences, University of Southampton, Highfield, Southampton, SO17 1BJ, UK
[#]ghyuan@ntu.edu.sg; [*]nzheludev@ntu.edu.sg

## Abstract

Evanescent light can be localized at the nanoscale by resonant absorption in a plasmonic nanoparticle or taper or by transmission through a nanohole. However, a conventional lens cannot focus free-space light beyond half of the wavelength $\lambda$. Nevertheless, precisely tailored interference of multiple waves can form a hotspot in free space of arbitrarily small size known as superoscillation. Here, we report a new type of integrated metamaterial interferometry that allows for the first time mapping of fields with deep subwavelength resolution $\sim \lambda/100$. It reveals that electromagnetic field near the superoscillatory hotspot has many features similar to those found near resonant plasmonic nanoparticles or nanoholes: the hotspots are surrounded by nanoscale phase singularities ($\sim \lambda/50$ in size) and zones where the phase of the wave changes more than tenfold faster than in a standing wave. These areas with high local wavevectors are pinned to phase vortices and zones of energy backflow ($\sim \lambda/20$ in size) that contribute to tightening of the main focal spot size beyond the Abbe-Rayleigh limit. Our observations reveal the analogy between plasmonic nano-focusing of evanescent waves and superoscillatory nano-focusing of free-space waves, and prove the fundamental link between superoscillations and superfocusing offering new opportunities for nanoscale metrology and imaging.

# Introduction

In recent years plasmonics – coupled electromagnetic states of light and free electrons in metals – has become a dominant research direction in photonics. The main advantage of plasmonics is that it gives access to large wavevectors of evanescent fields near metallic nanostructures that facilitate miniaturization and enhanced light localization in numerous applications (including sensors, photovoltaics, light harvesting, data storage, spectroscopy, ultracompact electro-optical devices and optical interconnects) and underpins the functionalities of advanced photonic materials and metamaterials engineered at the nanoscale. Moreover, evanescent plasmonic fields are often highly structured and contain phase singularities, vortices and energy backflow zones. The main drawback of plasmonics is the resistive Joule losses in metals that lead to rapid dissipation of energy in the nanostructures. Considerable efforts have been devoted to searching for novel plasmonic materials with functionalities and losses improved beyond those offered by conventional plasmonic media where material characteristics can be better than those offered by conventional plasmonic metals such as gold and silver [1-3].

The main purpose of this paper is to bring to the attention of the growing nanophotonics research community that the attractive features of evanescent plasmonic fields – such as high localization and extremely rapid variations of fields, giant wavevectors, phase singularities, nanoscale vortices and energy backflows – can be constructed not only in the immediate vicinity of the metallic plasmonic nanostructure, but also in free space, and therefore in the absence of Joule losses. Such extreme features can be generated in free space by diffraction of light on purposely constructed masks. Our work falls into the rich and fertile domain of singular optics and is stimulated and informed by pioneering works of Michael Berry and other researchers who theoretically predicted and observed complex field patterns in free-space optics including vortices and knots [4-13], largely without references to the plasmonic analogy.

However, our work goes further and explores one of the most practically important and fundamentally challenging questions of optics – how to focus light into a small spot – and exposes a previously unnoticed analogy between focusing by plasmonic nanostructures and superoscillatory focusing in free space.

In plasmonics, light evanescently confined near nanoparticles or nanostructures with rich spatial spectrum can change very rapidly and possesses high frequencies in its spatial spectrum. For instance, a plasmonic wave propagating along the end of a metallic taper can be concentrated down to nanometre dimensions [14]. Both an opaque screen with a small hole (typically a few tens of nanometers in diameter) and a tapered optical fiber can localize light into the diameter of the exit hole even if the wavelength of light is much larger than the hole. Such nearly point-like sources of evanescent light are used in the high-resolution imaging technique known as scanning near-field optical microscopy (SNOM). It is also known that evanescent electromagnetic fields can form vortices of energy flow (represented by the Poynting vector) in the near-field of a screen [15,16], in the interfacial region under the condition of total internal reflection [11], and near resonant plasmonic [17-19] or dielectric nanoparticles [20] excited by a plane wave. However, the evanescent component of the field does not propagate into the free space and decays rapidly away from the nanostructures.

But can light be localized (focused) to a very small hotspot in the free space far away from structured media? For a forward-propagating plane electromagnetic wave at frequency $\omega$ in free space, the projections of the wavevector $|\mathbf{k}_0| = \omega/c$ on any given direction $i$ are band-limited to $k_i \in [-\omega/c, \omega/c]$, where $c$ is the speed of light. This band-limit of light's spatial spectrum is often understood to yield the "diffraction limit": light cannot be structured smaller than certain scale using interference (i.e. through the formation of a standing wave). The common wisdom here results in the Abbe-Rayleigh rule for focusing by a conventional lens, claiming that, in free space, no lens can concentrate light into a spot smaller than half the wavelength.

As a matter of fact, within a finite interval, band-limited functions can *locally* oscillate much faster than their highest Fourier component. This is known as the phenomenon of superoscillation [6,21,22]. As explained by Michael Berry, superoscillations are possible because in the Wigner representation the local Fourier transform can have both positive and negative values, which causes subtle cancellations in the Fourier integration over all of the function [23]. Examples of superoscillatory functions are given in [24]. Applied to optics, the existence of superoscillations implies that if a number of waves with wavevectors bandlimited to $[-\omega/c, \omega/c]$ interfere in free space, the *local* spatial spectrum could contain high value wavevector with projections outside of the $[-\omega/c, \omega/c]$ band. Here we use the definition of local wavevector as $\mathbf{k}_{local} = \nabla\varphi$, where $\varphi$ is phase of the electromagnetic field in the locality and $\nabla$ is the gradient operator. In fact, there is no fundamental limit on how big the local wavevector can be. As a result, the free-space optical field created by interference of several band-limited waves, for instance by diffraction of a plane wave on a structured mask, can have deeply sub-wavelength spatial features. Such arbitrarily small foci of electromagnetic energy exist in free space and are far away from any boundaries with nanostructures or scattering objects thus offering a very powerful opportunity for developing far-field label-free super-resolution non-algorithmic microscopies at harmless levels of intensity without impregnating them with luminescent materials [25], which gives a crucial advantage over other super-resolution imaging techniques such as stimulated emission depletion (STED) [26] and single-molecule localization methods (SMLM) [27,28] that require labelling object with luminescent materials and high illumination intensity.

Superoscillations have been extensively studied in far-field super-resolution optical focusing and imaging [29-40], signal processing [41], electron wave confinement [42], wavefunction localization of a single photon [43], light trapping [44], ultrashort pulse generation [45], abnormal eigenvalues in quantum weak measurement [46], optical speckles [47], to name a few. In spite of these intense interests in the subject, no one has yet reported direct experimental evidence of large local wavevectors of superoscillatory optical fields in free space that are predicted to be closely linked to superoscillations [6,22]. Observing these effects is much more complex than demonstrating sub-Abbe-Rayleigh intensity hotspots: it needs not only an efficient generator of superoscillatory optical fields, but also the capability to recover the phase $\varphi$ of the superoscillatory portion of the field hence the local wavevectors $\mathbf{k}_{local} = \nabla\varphi$.

In this work, we report the first direct experimental observations that "sub-diffraction" superoscillatory optical focus in free space is surrounded by zones where the magnitude of local wavevector exceeds $|\mathbf{k}_0| = \omega/c$ by many times. Experimental observation of giant wavevectors around sub-Abbe-Rayleigh intensity hotspots is a triumph for the superoscillation concept.

Moreover, we show that the formation of superoscillatory foci breaking the Abbe-Rayleigh limit is linked to and facilitated by the formation of energy backflow zones in free space, far away from the structured mask that created the complex superoscillatory field. This brings us neatly to a very interesting theoretical prediction recently made by Berry [48] who demonstrated that in the field created by the interference of multiple waves with positive wavevectors, the local phase gradient can sometimes be negative. Here we show that energy backflows play a crucial role in the formation of superoscillatory foci which are pinned to phase singularities and energy nano-vortices (see Fig.1) even far from any scattering objects.

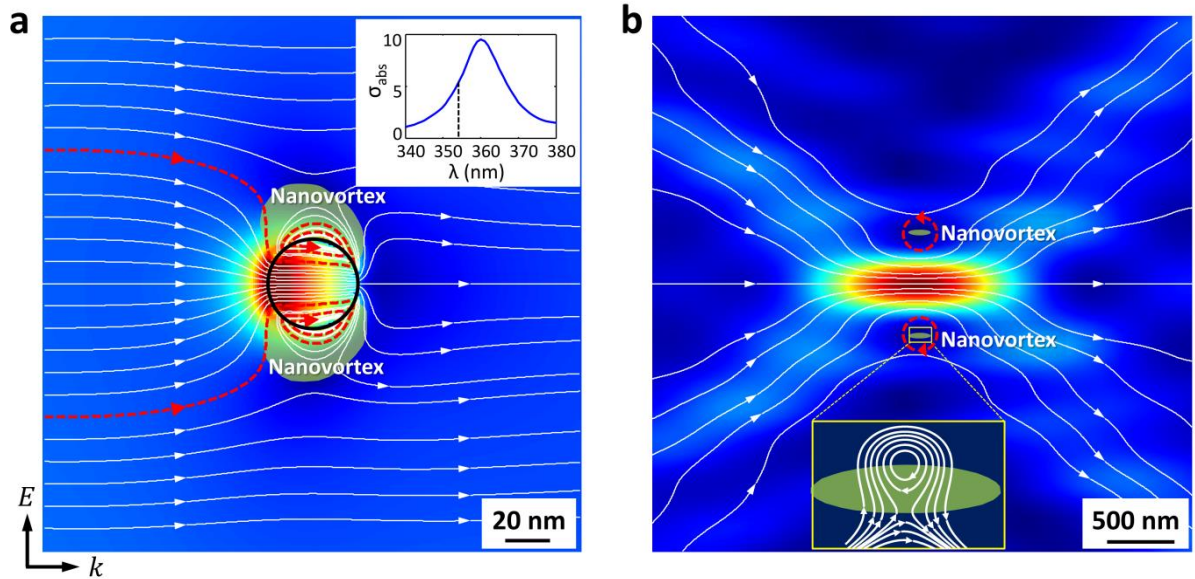

**Figure 1 | Focusing of light by a plasmonic nanoparticle and by a superoscillatory mask. a, Plasmonic nanoparticle (evanescent fields):** Poynting vector of light (white lines) near a plasmonic nanoparticle in resonance with incident field following Ref. [17]. Figure shows energy flow in the plane of polarization near a 40 nm-diameter silver nanoparticle illuminated at λ=354 nm, black circle indicates the nanoparticle boundary. Notice the evanescent field nano-vortices with energy backflow regions (shaded in green) at the edges of the nanoparticle. Inset shows the normalized absorption cross-section. **b, Superoscillatory focus (free-space fields):** Poynting vector of light near a superoscillatory focus in free space. In both panels the white lines show the direction of energy flow, and the red lines highlight the energy vortices. The color maps indicate the absolute value of the Poynting vector, with intensity increasing from blue to red. Notice nanovortices with energy backflow regions (shaded in green) near the superoscillatory focus.

In essence, our experimental observations reveal an intriguing similarity between the structures of evanescent fields focused by plasmonic nanostructures and that of the free-space superoscillatory focus. In both cases, giant local wavevectors, phase singularities, energy vortices and zones of backflow are present signifying and making possible the sub-wavelength localization of light (see Fig. 1).

## Nanoscale Interferometry of Superoscillatory Fields

There are two main challenges in providing complete mapping of superoscillatory fields. First, as phase information can only be recovered in interferometric measurements, extreme stability of the interferometer is needed to obtain reliable data on fields with small spatial features and fast phase variations. Second, the superoscillatory fields are expected to have spatial features that are much smaller than the wavelength of light: spatial resolution far better than allowed by the Abbe-Rayleigh limit of half-wavelength is required.

To meet these two challenges, we developed an original monolithic metamaterial interferometry (MMI). In this technique, the superoscillatory field under investigation and reference wavefront needed for interferometry are created by the same planar metamaterial nanostructure, i.e. on the same monolithic platform, thus minimizing the issues of stability and alignment which are characteristic of free-space interferometers [49-51].

Interference of the superoscillatory and reference fields creates a field distribution in free space that can be mapped with a linear detector array. Resolution of the detector array is determined by its pixel size and is insufficient for mapping. However, the field containing superoscillatory components and sub-wavelength features below the Abbe-Rayleigh diffraction limit is formed by the interference of free-space waves and therefore can be imaged with magnification by a lens with numerical aperture (NA) exceeding that of the diffraction mask to collect all the wavevector components. We used a CMOS camera array with pixel size of 6.5 μm together with a high magnification optics (500X), thus achieving a spatial resolution of 13 nm. It will be shown below that monolithic metamaterial interferometric technique allows detection of superoscillatory field features that are below 2% of the wavelength in size.

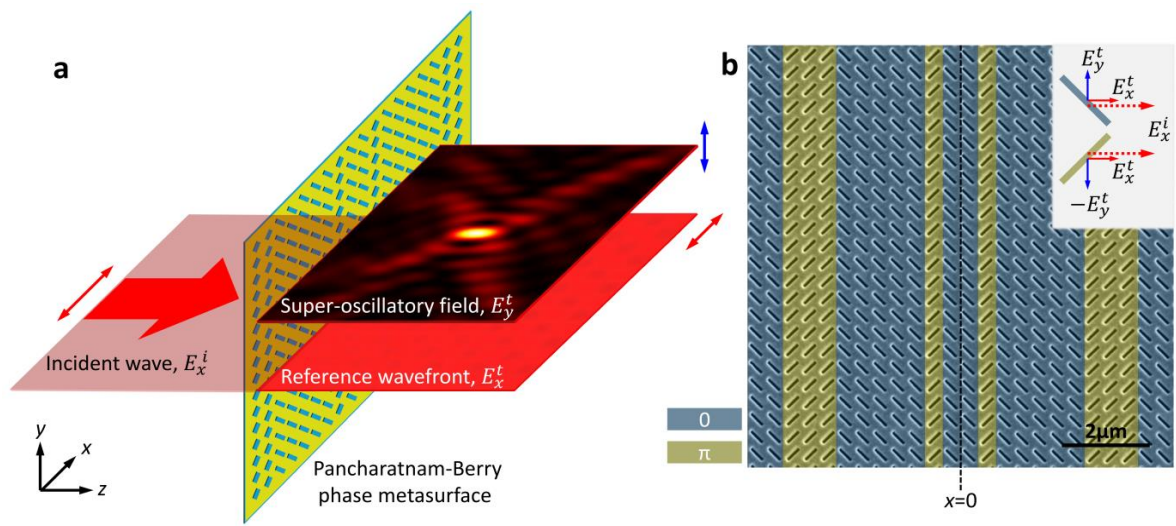

**Figure 2 | Monolithic metamaterial interferometry with a superoscillatory Pancharatnam-Berry phase metasurface. a**, Principle of monolithic metamaterial interferometry: a single metasurface creates the field distribution under investigation and the reference wave in orthogonal polarization. These two wavefronts are made to interfere on a polarization sensitive detector. **b**, Scanning electron microscopy image of the 40 μm by 40 μm metasurface fabricated in a 100 nm-thickness gold film with focused ion beam milling (unit cell size 400 nm by 400 nm) that acts as a superoscillatory field generator and interferometric platform. Artificial colors indicate rows providing binary $0/\pi$ phase shift in the transmitted light. The phase of light transmitted through the metasurface depends on the incident polarization and orientation of the slit, giving the opportunity to create a superoscillatory field for one of the transmitted polarizations and a plane wave for the orthogonal one, as explained in the text. A series of polarization-sensitive measurements of the intensity distribution of the diffracted light for different input polarizations allows unambiguous recovery of the phase map of the superoscillatory field.

The key component of the monolithic metamaterial interferometer is a metamaterial mask (Pancharatnam-Berry phase metasurface [52-54]) that simultaneously creates the tight superoscillatory focus and the reference wavefront for the interferometry, see Fig. 2a. It contains rows of identical scattering subwavelength slits oriented at either +45° or -45° with respect to the *x*-axis and has translation symmetry in *y* dimension thus working similarly to a cylindrical lens focusing light into a line focus. The working principle of the mask is shown in Fig. 2b. The mask is designed to be polarization-sensitive and creates a superoscillatory field only in cross-polarization to the incident wave. The field co-polarized with the incident wave will propagate through the mask as a plane wave – with some attenuation. For example when the metasurface is illuminated with *x*-polarized light ($E_x^i$ on Fig. 2b), the phase of *x*- polarized transmitted light ($E_x^t$) will be independent of the slit orientation: the *x*-polarized transmitted wave will remain a plane wave since the period is subwavelength and only the zero diffraction order is generated. However, slits oriented at +45 ° and -45 ° to *x*-axis will transmit *y*-polarized light with $\pi$ phase difference ($E_y^t$ and $-E_y^t$, see Fig. 2b). Therefore, the pattern of slits works as a binary phase grating for *y*-polarization. We define this arrangement as the TE configuration. Similarly, when illuminated with *y*-polarized light, the metasurface is a binary phase mask for transmitted *x*-polarization (TM configuration).

Such a mask allows for a straightforward interferometry between the superoscillatory field and the reference plane wavefront that are mutually stable and inherently aligned by design. A 3D map of intensity and phase can be recorded by measuring intensity distributions at different distances from the mask using different input polarizations and polarization-sensitive detection.

Upon transmission through the metasurface the *x*-polarized field suffers the same phase retardation *irrespective of* the orientation of the slits and the same intensity attenuation at all points due to the energy transfer into the cross-polarised field. Therefore, for the *x*-polarized field the metasurface is a *homogeneous subwavelength grating* of limited size (aperture). It will produce only a zero-order diffraction field which does not depend on the state of polarization of light incident on the metasurface. Although the *x*-polarized field shows some variations from the plane wave due to aperture diffraction at the edges of the metasurface, it is a good reference field for interferometry as it has phase close to that of a plane wave (see Supplementary Information Section 3) and a well-defined, easy-to-measure intensity profile with no zeros.

In our experiment, the metasurface contains 100 rows of slits. The metasurface grating is designed with the particle swarm optimization algorithm [55] to generate superoscillatory foci of prescribed spot size, focal distance, field of view and depth of focus (see Supplementary Information – Section 1 for the metasurface design details). Indeed, examples given in Refs. [24,25] illustrate that interference of only a few waves is sufficient to generate a superoscillatory field. Since the grating creates a superoscillatory focus for one polarization only while the transmitted light remains a plane wave for the other, the phase distributions in the TE and TM configurations $\varphi_{TE,TM} = \arg(E_{y,x})$ can be mapped by measuring the intensity distribution $I(x,z)$ of the interference pattern at a distance *z* from the mask for different polarizations of incident light (*x*, *y*, +45 °, -45 °, right and left circular) as follows:

$$\varphi_{TE} = \operatorname{atan}\left(\frac{I_y^{LCP}-I_y^{RCP}}{I_y^{+45°}-I_y^{-45°}}\right) + k_0 z \quad (1)$$

$$\varphi_{TM} = \operatorname{atan}\left(\frac{I_x^{RCP}-I_x^{LCP}}{I_x^{+45°}-I_x^{-45°}}\right) + k_0 z \quad (2)$$

Here the second term on the right hand-side of equations (1) and (2) comes from the reference plane wave, the superscripts and subscripts denote polarization of the incident light and detection light correspondingly (see Supplementary Information – Section 2 for the derivation details).

Results of mapping the interference patterns $I(x,z)$ for TE configuration can be found in Fig. 3. Figure 3a shows performance of the metasurface evaluated by finite-difference time-domain (FDTD) calculation. When incident light at the wavelength of 800 nm is *x*-polarized, the *y*-polarized component of the diffracted wave contains a superoscillatory hotspot at 10 µm away from the metasurface. The focal spot has a full-width at half-maximum (FWHM) of

0.42$\lambda$, which is well below the Abbe-Rayleigh diffraction limit $\lambda$/2NA=0.56$\lambda$ for a cylindrical lens with numerical aperture corresponding to the experimental situation of NA=0.89 (20 μm wide lens with a focal distance of 10 μm). Under *y*-polarized illumination, the *y*-polarized component of the diffracted wave is the reference field that we use for interferometry. For an infinitely long grating, it would show no structural features, while minor variations in the transmission amplitude observed experimentally are due to the aperture effects. Its phase is uniform and therefore intensity variations do not affect accuracy of the phase retrieval (see Supplementary Information – Section 3). With circularly polarized incident waves, the diffraction patterns $I_y^{RCP,LCP}(x,z)$ originate from the interference of the superoscillatory field and the reference wave with an initial phase difference of ±$\pi$/2 between them, depending on the handedness of the incident polarization. Similarly, for the incident linear polarization at ±45°, when we measure $I_y^{\pm 45°}(x,z)$, the phase difference between the superoscillatory and reference fields becomes 0 and $\pi$ respectively. Now the phase map $\varphi_{TE}(x,z)$ of the superoscillatory field can be recovered from $I_y^{RCP,LCP}(x,z)$ and $I_y^{\pm 45°}(x,z)$ maps using formula (1).

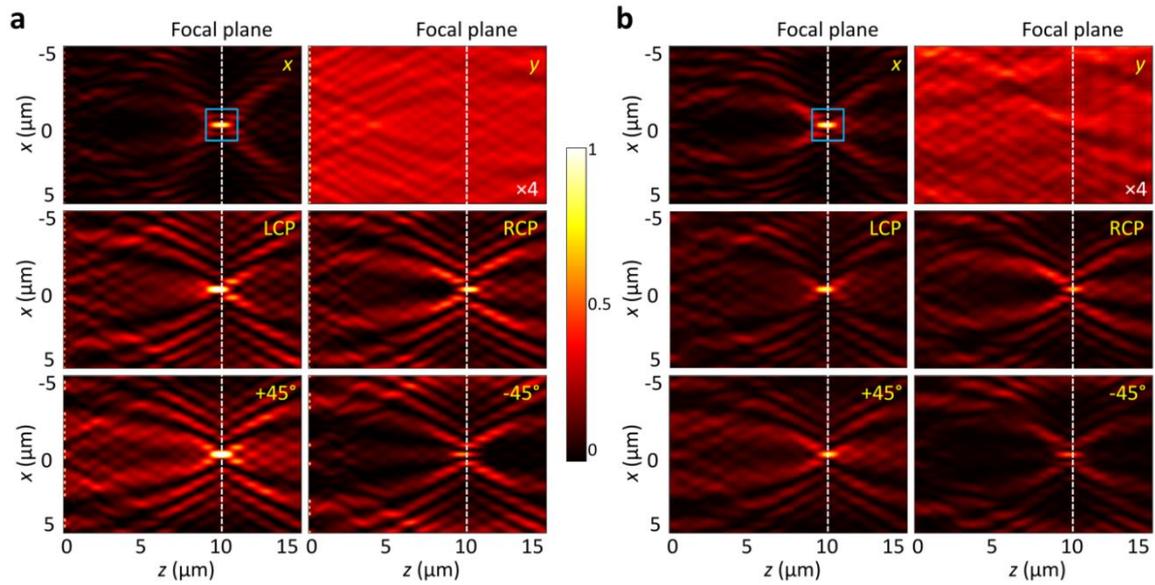

**Figure 3 | Superoscillatory field generated by the metasurface under different illumination conditions. a**, FDTD simulation; **b**, Experimental data. Figures show the *x-z* cross-section of the *y*-component intensity distribution in the interference pattern $I_y(x,z)$ for different incident polarizations (indicated in the corners of the maps – TE configuration). Note that the formation of superoscillatory hotspot (highlighted in the blue box) located at distance of 10 μm from the metasurface is best manifested for orthogonal incident polarization ($E_x$). The quasi-uniform field map for the parallel incident polarization ($E_y$) is multiplied by a factor of 4 and behaves like a reference plane wave for creating the interferogram with the signal superoscillatory field. FDTD computed and experimentally measured maps show good agreement.

In our experiment, we used an 800 nm wavelength diode laser as optical source and mapped the interference pattern $I(x,z)$ with a CMOS camera placed on a nanometric translation stage and equipped with a 500X magnification optical system. See Supplementary Information – Section 4 for the detailed optical characterization process. Corresponding experimental results for the intensity maps are given in Fig. 3b. Good agreement with the calculated maps is found in all field patterns. The co-polarization light shows intensity variations due to aperture diffraction on the mask edges. Some asymmetry in the pattern in Fig. 3b is due to imperfections in the incident wavefront. From quantitative calculation, the mean square difference between a plane wave wavefront and simulated and measured field maps was found to be 11.7% and 24.1% respectively. Here the superoscillatory hotspot (upper left panel in Fig. 3b) is also observed at $z=10$ μm and its FWHM is measured to be $0.43\lambda$, only about 2% difference from the computed size of the hotspot.

## Four Features of Superoscillatory Fields

Figure 4a shows computed and measured *x-z* cross-sections of the intensity map in the near vicinity of the superoscillatory focus, annotated by the blue box in Fig. 3. Here we can see the *first characteristic feature of superoscillatory optical field, high localization of the field*. The hotspot size in the *x*-section is smaller than allowed by the Abbe-Rayleigh limit (because it does not take account of superoscillation). The focus is surrounded by fringes similarly to how focal spot of a conventional lens of finite size is surrounded by the oscillating Airy pattern. However, here the fringes are more densely spaced than in the Airy pattern and much more extensive fringes are present. Indeed, superoscillatory hotspots are always surrounded by intense halos or fringes [24]. At the focal plane, the intensity of the first sidelobe is 17.6% (simulation) and 16% (experiment) of the peak intensity of the central hotspot.

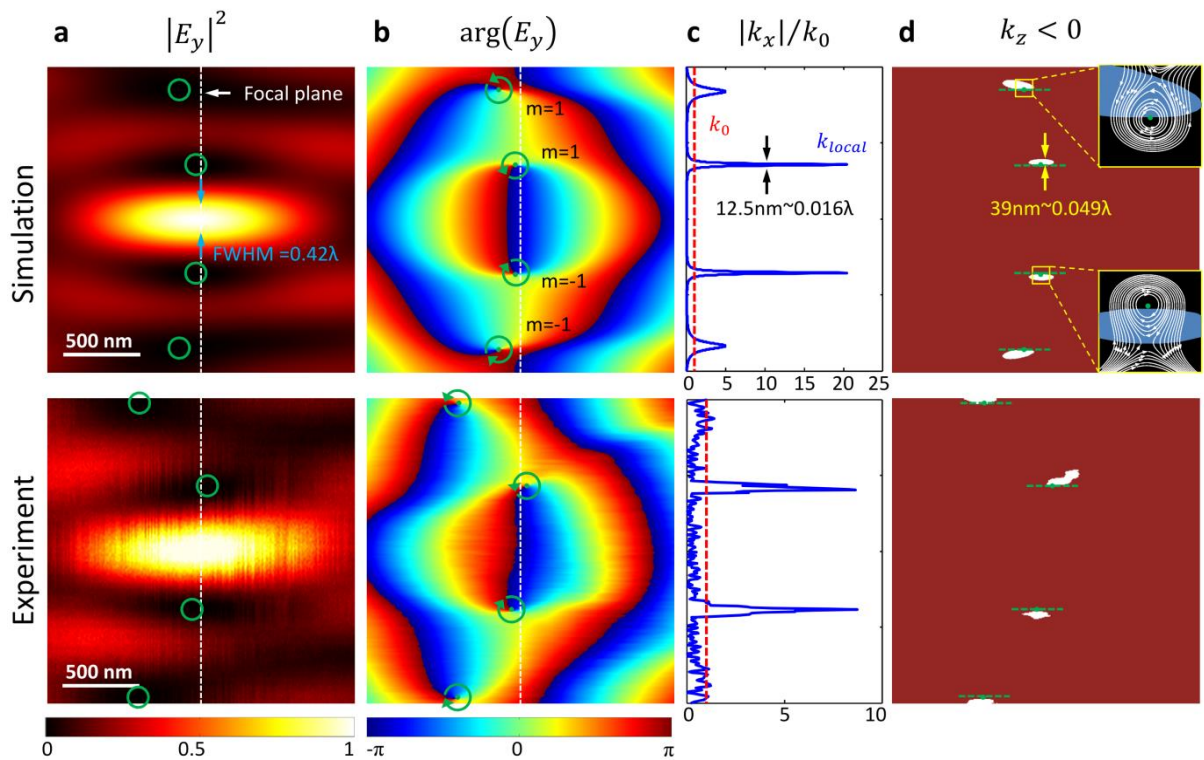

**Figure 4 | Four characteristic features of a superoscillatory field.** (Top row) simulation; (Bottom row) experiment. **a, High localization of the field** can be seen from the intensity map $|E_y|^2$ depicting focus with FWHM of $0.42\lambda$, smaller than allowed by the Abbe-Rayleigh limit. The vertical dashed line indicates plane of the focus; **b, Phase singularities** with topological charge of $m=\pm 1$ (green dots) are seen on the phase maps $\arg(E_y)$ in the low-intensity areas of the superoscillatory field; **c, Gigantic local wavevectors** $|k_x|/k_0$ at the focal plane are calculated from the phase gradient in *x*-direction. Superoscillatory values of local wavevectors are highly localized in zones of order of $\lambda/100$; **d, The energy backflow (retro-propagation)** areas are painted white. They are substantially sub-wavelength in size (~$\lambda/20$ along *x*-direction) and correspond to negative values of $k_z$ calculated from the phase gradient in *z*-direction. Insets with the black background show zoom-in view of the Poynting vectors near the phase singularity, and backflow regions are shaded in blue. Dashed green lines indicate tangent to the retro-propagation areas at the point of their intersections with phase singularities.

The phase $\varphi_{TE}(x,z)$ of the electromagnetic field is rapidly changing near the superoscillatory focus, as shown in Fig. 4b. Here one can observe a close match between computed and experimentally measured phase maps that were retrieved from intensity maps in Fig. 3b using formula (1). On the phase maps, one can clearly observe *the second characteristic feature of superoscillatory optical fields*: they are accompanied by *phase singularities*. At the low-intensity regions near the focus, one can see four phase singularities identified by green circles (Fig. 4(b)). When moving along the loop encircling the phase singular points, the phase changes by $2\pi$. The two singularities in the upper part of the phase map have topological charge of $m=+1$, while in the lower part they have topological charge of $m=-1$.

As *the third characteristic feature of superoscillatory optical field*, we observed *gigantic local wavevectors* in the field maps with values far exceeding $k_0 = \omega/c$. First, we calculated the transverse wavevector $k_x$ as $x$-component of the gradient of the computed and measured phase values. In the experiment, the pixilation of the phase mapping was 13 nm along the $x$-direction and 10 nm along the $z$-direction. The normalized transverse local wavevector $|k_x|/k_0$ at $z=10$ μm is given in Fig. 4c. From there we see that the $|k_x|$ near the phase singularity is more than an order of magnitude higher than $k_0 = \omega/c$. Here computed and experimental data are in good agreement qualitatively, although somewhat smaller in the experiment but still very large wave-numbers beyond the spectrum are observed. Here we note that phase retrieval and wavevector mapping are robust to noise, see Supplementary Information – Section 5.

The presence of *the fourth characteristic feature of superoscillatory optical field*: the existence of *the energy backflow (retro-propagation)* areas near the superoscillatory focus, can be derived from the mapping of longitudinal wavevector $k_z$. On the $x$-$z$ maps we painted the areas of energy backflow as white zones in Fig. 4d. Indeed, we observed that near the phase singularities $k_z$ can have negative values. Since the Poynting vector is always parallel to the local wavevector, negative values of $k_z$ mean energy back-flow (see insets presenting detailed computed Poynting vector maps near phase singularities). As can be clearly seen in the inset of Fig. 4d, phase singularities are pinned to the energy backflow regions, as predicted in [48]. Here one can see that the incident energy flow is "trapped" and circulates without propagating in the forward direction (compare with energy flow near the plasmonic nanostructure, see Fig.1). As predicted by Berry for a general case of interfering multiple waves [48], "The boundaries of the retro-propagating regions include the phase singularities … and are tangent to the $z$ direction at these points" (see green dashed lines on Fig. 4d. They are consistently directed along $z$ which confirms accuracy of the experiment). Here we also confirm another powerful observation from the same paper that "the regions of backflow are considerably smaller than the wavelength; this reflects the well-known fact that in the neighbourhood of phase singularities wavefunctions can vary on sub-wavelength scales." Indeed, backflow areas are only about $\lambda/20$ in size along $x$-direction, but are still satisfactorily resolved and their positions are accurately mapped on computed locations. It is also noteworthy that due to translation symmetry no energy backflow is observed in the plane perpendicular to the $x$-$z$ plane.

The existence of phase singularities and energy backflow zones pinned to optical superoscillations gives a qualitative insight into the mechanism of focusing beyond the Abbe-Rayleigh limit. Two singularities that are close to the superoscillatory focus are located in the areas of diminishing intensity that define the boundary of the focus. At the superoscillatory focus, the backflow depletes the area where flow propagates in the forward direction thus narrowing the focus beyond the conventional diffraction limit. This could be compared with laminar energy flow at the Gaussian beam focus where energy propagates in the forward direction only (see Supplementary Information – Section 6).

To emphasize the great degree of similarity between plasmonics and superoscillatory free-space fields, we compare the main four features of the super-oscillatory field described above with the iconic case of the fields at plasmonic nanoparticle (see Fig. 5). *The first characteristic feature of superoscillatory optical field, high localization of the field*, is a well-known feature of plasmonic resonances that is well illustrated by Fig. 5a. *The second characteristic feature of superoscillatory optical fields, phase singularities,* can be seen from Fig. 5b. As in free-space superoscillatory fields, singularities are pinned to area of low intensity, see from the inset in Fig. 5a where the intensity near the interface shows a minimum. *The third characteristic feature of superoscillatory optical field*, *gigantic local wavevectors, can also be seen in plasmonics*, as shown in Fig. 5d. Moreover, as in free space superoscillatory fields, the large wavevectors $|k_x|>k_0$ (white areas on Fig. 5d) exist in the low-intensity regions (deep blue areas in Fig. 5a). *The fourth characteristic feature of superoscillatory optical field*: *the energy backflow (retro-propagation)* can also be seen in plasmonic field maps in Fig. 5c, which also exhibits another important feature of the free-space superoscillatory field observed in our work: the boundaries of the retro-propagating regions include the phase singularities (compare Fig. 5b and Fig. 5c).

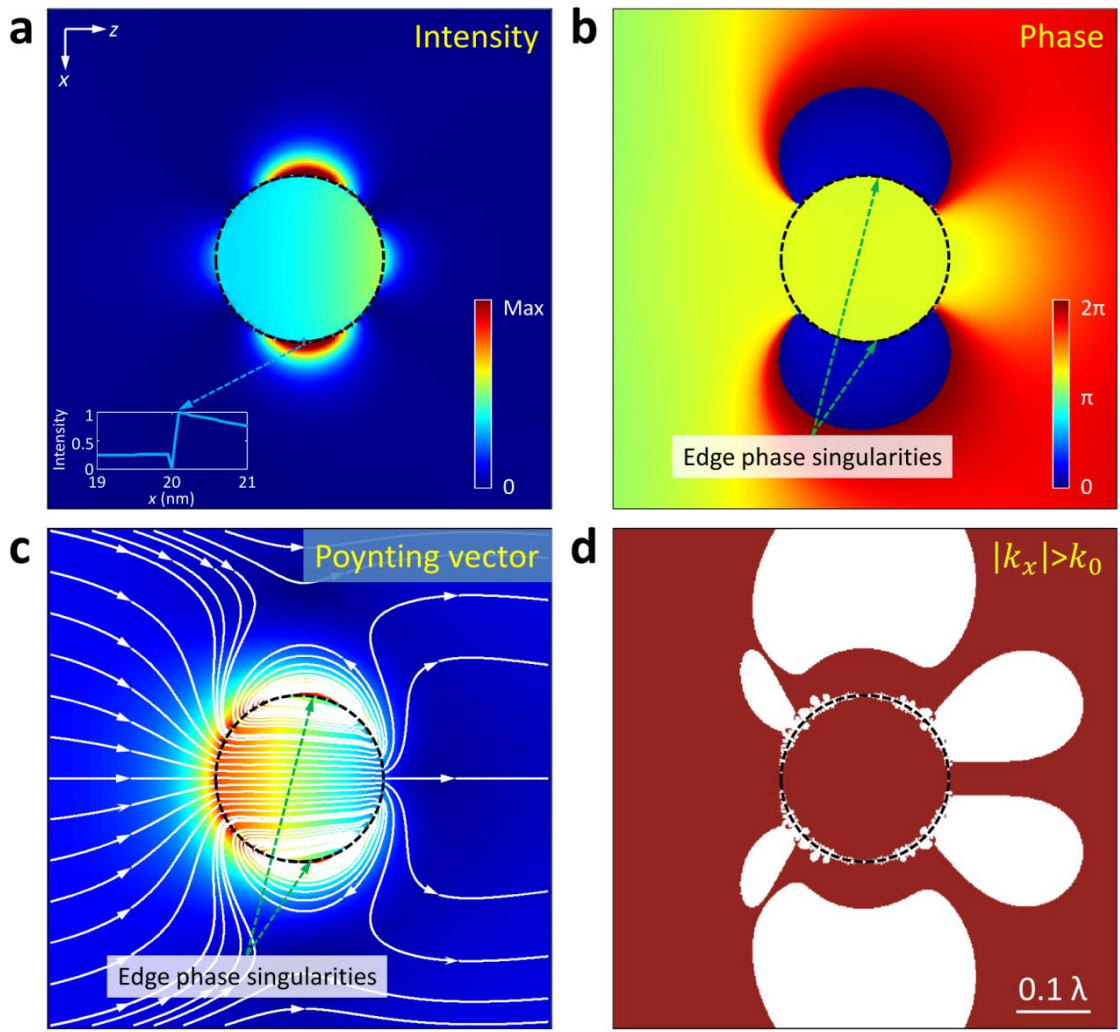

**Figure 5 | High localization, phase singularities and energy backflow in plasmonic fields. a,** Intensity, **b,** phase of the electric field near plasmonic resonance. The line field map across the boundary in (a) is shown in the inset where the low intensity region is clearly seen. **c,** The Poynting vector map; **d,** Large wavevector regions, areas with $|k_x|>k_0$ are coloured in white. All parameters are the same as for Fig. 1a. Black dashed circles show the boundary of the silver nanoparticle. The phase singularities are highlighted by the green dashed lines in (b) and (c).

Similar results for the TM configuration are presented in Supplementary Information – Section 7 where all the four characteristic features of superoscillatory fields including the high localization of field, phase singularities, gigantic local wavevectors and energy backflow are also experimentally observed.

## Conclusion

Using new monolithic metamaterial interferometry, we comprehensively mapped a two-dimensional superoscillatory electromagnetic field generated by a Pancharatnam-Berry phase metasurface with unprecedented resolution of about λ/100. We have been able to experimentally prove and simultaneously observe, for the first time, four characteristic features of optical superoscillations: sub-Abbe-Rayleigh localization of light in the focus, phase singularities and nano-vortices, gigantic local wavevectors and the energy backflow (retro-propagation) in the immediate proximity of the focus. Our result will be useful for better understanding of superoscillatory focusing in wave optics [33-40] and electron waves [42] and imaging applications [32,36,39].

Our observations have identified some remarkable similarities between near-field plasmonic focusing by nanostructures and superoscillatory focusing in free space. Indeed, in both cases, giant local wavevectors and phase singularities are observed surrounding the plasmonic and superoscillatory foci; energy vortices and zones of backflow are pinned to phase singularities and facilitate the sub-wavelength localization of light by removing electromagnetic energy from the areas neighbouring the foci. See Fig. 1.

Here we shall note that there is another important common feature of superoscillatory and near-field plasmonic focusing that does not directly follow from our observations, but rather represents basic scaling laws: the efficiency of focusing scales polynomially with size of the focal spot. Indeed, a deeply sub-wavelength hole of diameter $\sigma$ in opaque screen can be used as a nanoscale light source, for instance in scanning near-field optical microscopy. Only a small proportion of light illuminating the screen will pass through the hole. Here throughput efficiency scales as $(\sigma/\lambda)^4 + o(\sigma/\lambda)^4$ [56]. A small absorbing dielectric or plasmonic nanoparticle of diameter $\sigma$ will also "focus" light by harvesting energy of the illuminating wave (see Fig. 1a) with scattering cross-section that scales as $(\sigma/\lambda)^4$ [57]. Similarly, in the regime of superoscillatory focusing in free space, only a small fraction of light can be focused in the hotspot. From the general theory of superoscillations [41], it follows that the proportion of energy channeled into the superoscillatory region decreases polynomially $P(\sigma/\lambda)$ with size of the superoscillation.

Finally, we have shown that monolithic metamaterial based interferometry allows robust generation of extremely small (~λ/100) spatial features surrounding superoscillations such as energy backflow zones and nano-vortices by a single nanostructured metasurface. This offers interesting opportunities for metrology, for instance for measuring lateral mutual displacement of two platforms when one platform contains the light source and the metasurface and the other an imaging light detector identifying position of the sub-wavelength features projected on it by the metasurface. We shall provide detailed description of such metrology application in a separate paper.

## Acknowledgments

This work was supported by the Singapore Ministry of Education (Grant MOE2011-T3-1-005), ASTAR QTE Programme Grant SERC A1685b0005, the Engineering and Physical Sciences Research Council UK (Grant EP/G060363/1). The data from this paper can be obtained from the University of Southampton ePrints research repository: http://doi.org/10.5258/SOTON/D0378.

## Author contributions

NIZ and GHY conceived the idea of phase measurement of optical superoscillatory field with metasurface. GHY and ETFR carried out the superoscillatory mask design and simulations. GHY performed the fabrication, measurement and data analysis. NIZ and GHY wrote the paper. All authors discussed the results and their interpretation and cross-edited the manuscript. NIZ supervised and coordinated all the works.

## Competing financial interests

The authors declare no competing financial interests.

# Supplementary Information for

# "Plasmonics" in free space: observation of giant wavevectors, vortices and energy backflow in superoscillatory optical fields

**Guang Hui Yuan[1#], Edward T. F. Rogers[2,3], and Nikolay I. Zheludev[1,2*]**

[1]*Centre for Disruptive Photonic Technologies, TPI, SPMS, Nanyang Technological University, Singapore 637371, Singapore*

[2]*Optoelectronics Research Centre and Centre for Photonic Metamaterials, University of Southampton, Highfield, Southampton, SO17 1BJ, UK*

[3]*Institute for Life Sciences, University of Southampton, Highfield, Southampton, SO17 1BJ, UK*

[#]ghyuan@ntu.edu.sg; [*]nzheludev@ntu.edu.sg

## 1. Metasurface design

The superoscillatory Pancharatnam-Berry phase metasurface is designed using the binary particle swarm optimization (BPSO) algorithm. The two dimensional metasurface at the *xy* plane is divided into *N* =50 pair of equally spaced rows of slits (Δ*x*=400 nm) placed with mirror symmetry with respect to *x*=0 plane. Each row has slits oriented either at +45 °or -45 ° with respect to the *x*-axis and provides phase delay of either $\pi$ or 0 respectively. The BPSO algorithm optimizes the light field distribution created by the metasurface near the focal plane when illuminated with plane wave. The merit function to be optimized is defined as:

$$I^{tar}(x,z) = [\text{sinc}(ax)]^2 \exp\left[-\frac{(z-z_f)^2}{b^2}\right] \qquad (1)$$

where $z_f$ is the desired focal length, $a = \frac{0.886}{\sigma}$, $b = \frac{D}{2\sqrt{ln2}}$, σ is the full-width at half-maximum of the hotspot size and *D* is the depth of focus.

## 2. Operating principle of the phase retrieval technique

The metamaterial mask produces a superoscillatory field in $E_y^x$ and a plane wave in $E_x^x$ under *x*-polarized excitation (see Fig. 2(c)), and a superoscillatory field in $E_x^y$ and a plane wave in $E_y^y$ under *y*-polarized excitation. Note that the superscripts and subscripts denote polarization of the incident field and detected field respectively. If illuminated by left-handed circularly polarized ('LCP') light, the transmitted field can be expressed as

$$\begin{bmatrix} E_x^{LCP} \\ E_y^{LCP} \end{bmatrix} = \frac{1}{\sqrt{2}} \begin{bmatrix} E_x^x + iE_x^y \\ E_y^x + iE_y^y \end{bmatrix} \qquad (2)$$

And the corresponding intensity is given by

$$\begin{bmatrix} I_x^{LCP} \\ I_y^{LCP} \end{bmatrix} = \begin{bmatrix} (I_x^x + I_x^y)/2 - \sqrt{I_x^x \cdot I_x^y} \sin\delta_{TM} \\ (I_y^x + I_y^y)/2 + \sqrt{I_y^x \cdot I_y^y} \sin\delta_{TE} \end{bmatrix} \quad (3)$$

where $\delta_{TM} = \varphi_{TM} - \varphi_{PW}$ is the phase difference between the superoscillatory field $E_x^y$ ($\varphi_{TM} = \arg(E_x^y)$) and the plane wave $E_x^x$ ($\varphi_{PW} = k_0 z$), while $\delta_{TE} = \varphi_{TE} - \varphi_{PW}$ is the phase difference between the superoscillatory field $E_y^x$ ($\varphi_{TE} = \arg(E_y^x)$) and the plane wave $E_y^y$ ($\varphi_{PW} = k_0 z$).

Similarly, for right-handed circularly polarized light ('RCP') and ±45° linearly polarized light

$$\begin{bmatrix} I_x^{RCP} \\ I_y^{RCP} \end{bmatrix} = \begin{bmatrix} (I_x^x + I_x^y)/2 + \sqrt{I_x^x \cdot I_x^y} \sin\delta_{TM} \\ (I_y^x + I_y^y)/2 - \sqrt{I_y^x \cdot I_y^y} \sin\delta_{TE} \end{bmatrix} \quad (4)$$

$$\begin{bmatrix} I_x^{\pm 45°} \\ I_y^{\pm 45°} \end{bmatrix} = \begin{bmatrix} (I_x^x + I_x^y)/2 \pm \sqrt{I_x^x \cdot I_x^y} \cos\delta_{TM} \\ (I_y^x + I_y^y)/2 \pm \sqrt{I_y^x \cdot I_y^y} \cos\delta_{TE} \end{bmatrix} \quad (5)$$

From equations (3)-(5), we derive that:

$$\tan(\delta_{TM}) = \frac{I_x^{RCP} - I_x^{LCP}}{I_x^{+45°} - I_x^{-45°}} \quad (6)$$

$$\tan(\delta_{TE}) = \frac{I_y^{LCP} - I_y^{RCP}}{I_y^{+45°} - I_y^{-45°}} \quad (7)$$

And therefore,

$$\varphi_{TM} = \operatorname{atan}\left(\frac{I_x^{RCP} - I_x^{LCP}}{I_x^{+45°} - I_x^{-45°}}\right) + k_0 z \quad (8)$$

$$\varphi_{TE} = \operatorname{atan}\left(\frac{I_y^{LCP} - I_y^{RCP}}{I_y^{+45°} - I_y^{-45°}}\right) + k_0 z \quad (9)$$

Therefore, by measuring intensity maps ($I_x^{LCP}$, $I_x^{RCP}$, $I_x^{+45°}$, $I_x^{-45°}$) and ($I_y^{LCP}$, $I_y^{RCP}$, $I_y^{+45°}$, $I_y^{-45°}$), we are able to retrieve the phase of the superoscillatory field in TM and TE configuration respectively. The transverse and longitudinal local wavevectors can then be calculated from $k_x = \nabla_x \varphi$ and $k_z = \nabla_z \varphi$.

## 3. Reference wave

To assess quality of the reference wave used in the interferometry, we simulated the field structure created by the metamaterial array using a finite-difference time-domain (FDTD) Maxwell equation solver (Lumerical, FDTD Solutions) for the TE configuration. The results are shown in Fig. S1(a). The reference wave $E_x^x$ shows a plane wavefront, although there is a slight intensity modulation due to the aperture effects and pixilation of the metasurface. Similar results are obtained for the TM configuration, confirming a flat wavefront for the reference wave $E_y^y$ (Fig. S1 (b)).

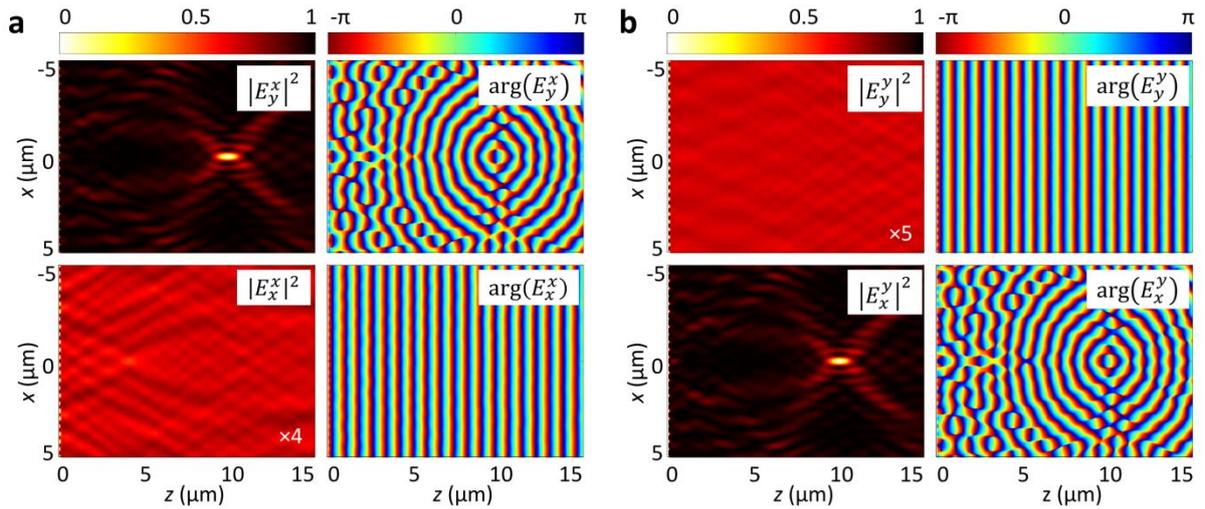

**Figure S1.** FDTD simulated phase profiles of the superoscillatory field and reference wave. A plane wavefront could be observed in the reference waves although the intensity distributions have slight variations due to aperture effects. **a**, TE configuration; **b**, TM configuration.

## 4. Optical characterization

The experimental setup is sketched in Fig. S2. The laser source is a diode laser with emission wavelength of 800 nm (Toptica DLC DL pro 780) and linewidth of 100 kHz. A linear polarizer (P1) and a quarter waveplate (QWP) or half waveplate (HWP) are used to produce the desired incident polarization. The field distribution created by the metamaterial mask was mapped with a high-resolution camera (Andor Neo sCMOS, 2560*2160 pixels, pixel size 6.5 μm) and a high-magnification objective (Nikon CFI LU Plan APO EPI 150X, NA=0.95) with a 4X magnifier. The actual magnification factor is calibrated to be ~500, corresponding to an effective pixel size of 13 nm. Another polarizer (P2) is inserted into the optical path before the camera to select the desired detection polarization state ($E_y$ in the TE case, $E_x$ in the TM case). The field maps were obtained by z-scanning with a step size down to 10 nm using a piezo stage (PI E517).

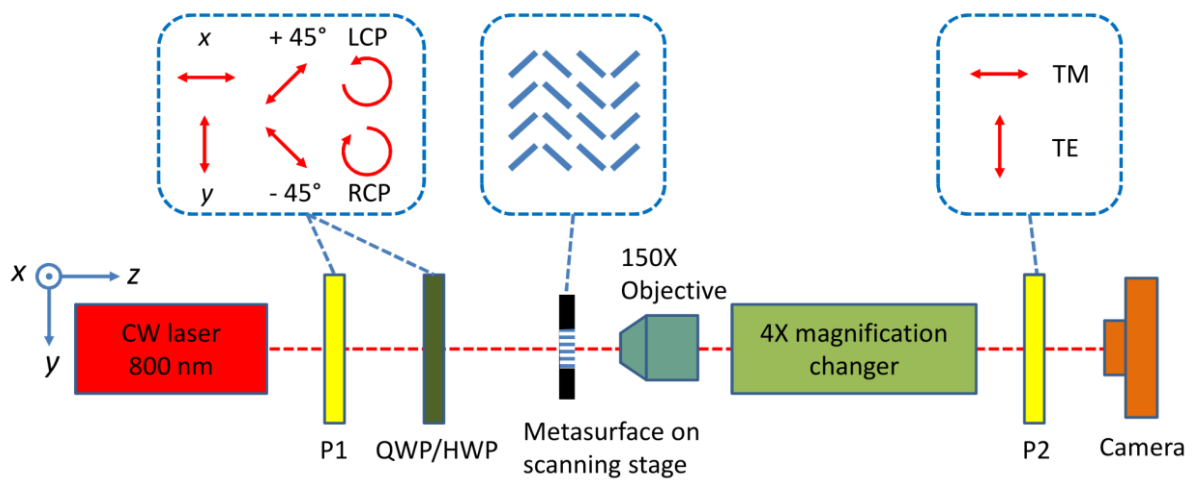

**Figure S2.** Experimental setup for generation and phase measurement of superoscillatory field using a metasurface. The superoscillatory field and reference plane wave are inherently aligned and coaxially propagating. P1, P2: linear polarizer; QWP: quarter waveplate; HWP: half waveplate. The top three panels show the z-view of the six incident polarizations (left) and two detection polarizations (right) used for intensity profile measurement and the schematic arrangement of the metasurface (middle).

# 5. Sensitivity of the phase retrieval to noise

To address the question of stability of phase recovery from intensity maps using formula (1-2) of the main text, we simulated the impact of adding white noise to the phase retrieval process. The results are summarized in Fig. S3. Here the noise intensity level is defined as the ratio of the intensity fluctuation range to the maximum intensity of the focal spot at $z=10$ μm. In absence of noise, the phase and local wavevector are presented in Fig. S3(a). Fig. S3(b) shows the same maps at the noise level of 20%. All main features of the phase portrait are clearly seen, indicating excellent robustness of the phase recovery process to laser intensity fluctuations and noise in the CDD detector array.

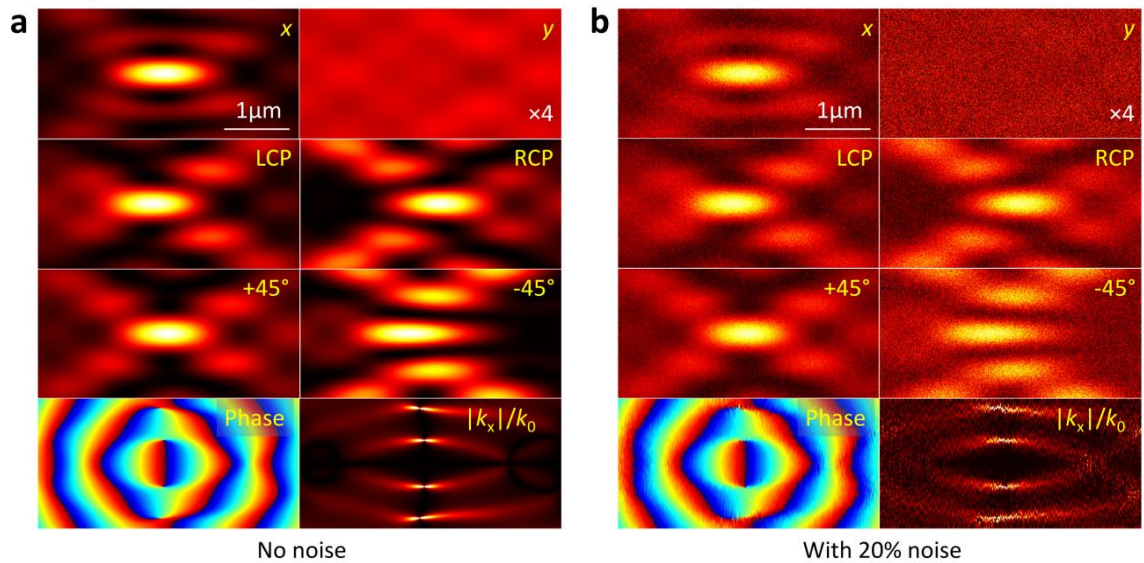

**Figure S3.** The intensity distributions and retrieved phase and local wavevectors: (a) without noise; (b) with 20% noise.

# 6. Energy flow near a conventional Gaussian beam focus

A conventional linearly polarized Gaussian beam in paraxial approximation can be expressed as $E(r,z) = E_0 \frac{w_0}{w(z)} \exp\left[-\frac{r^2}{w(z)^2}\right] \exp\left[-i\left(k_0 z + \frac{k_0 r^2}{2R(z)} - \psi(z)\right)\right]$, where $w_0$ is the beam waist, $w(z) = w_0\sqrt{1+\left(\frac{z}{z_R}\right)^2}$ is the beam size, $z_R = \frac{\pi w_0^2}{\lambda}$ is the Rayleigh distance, $R(z) = z\left[1+\left(\frac{z_R}{z}\right)^2\right]$ is the curvature radius of the wavefront, $\psi(z) = \operatorname{atan}\left(\frac{z}{z_R}\right)$ is the Gouy phase. Figure S4 gives an example of the Poynting vector distributions calculated from the local wavevectors. No energy backflow phenomenon is observed near the focus. It should be noted that tightly focused beams from lenses with limited aperture and aberration will often exhibit phase singularities and weak energy backflow near the focal plane.

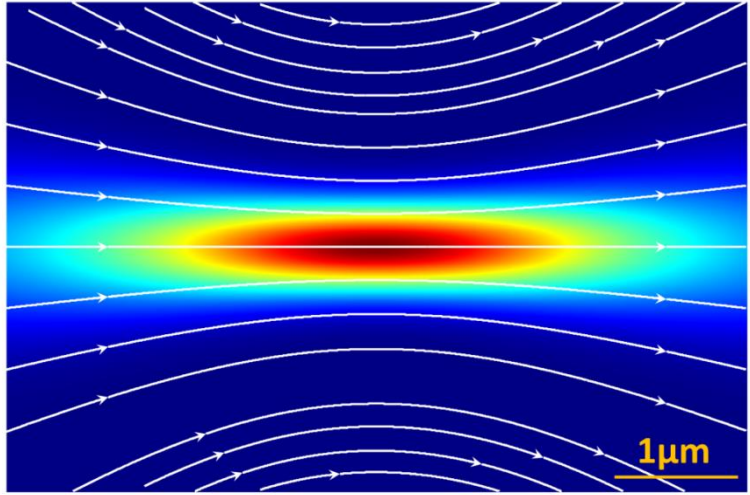

**Figure S4.** Poynting vector near a conventional Gaussian beam focus. The wavelength is λ=800 nm, $w_0$=1μm.

## 7. Transverse magnetic (TM) configuration

In the main manuscript, we present results for the TE configuration. For completeness, we present equivalent field maps for the TM configuration in Figs. S5 & S6. The superoscillatory fields are created in $E_x$ under *y*-polarized excitation with calculated and experimentally achieved hotspot of $0.41\lambda$ and $0.42\lambda$ respectively. Four superoscillatory regions with $|k_x| > k_0$ are clearly observed in Fig. S6(c), and the energy backflow regions with $k_z < 0$ highlighted in white are shown in Fig. S6(d).

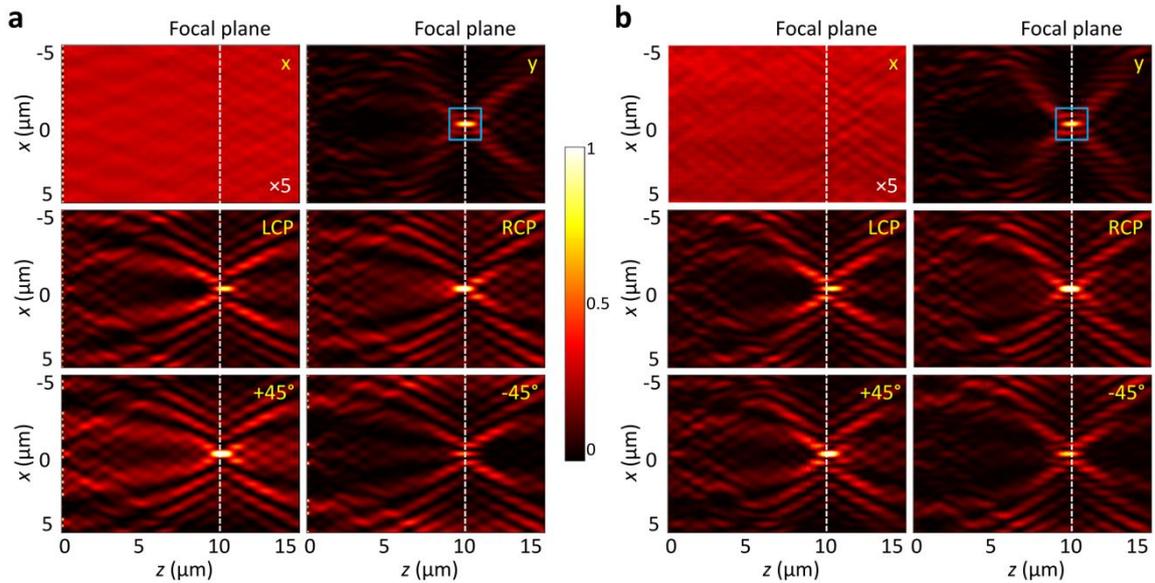

**Figure S5.** Superoscillatory field generated by metasurface at different illumination conditions (TM case). **a**, FDTD simulation; **b**, Experimental data. The superoscillatory focus located at a propagation distance of 10 μm, as annotated by the blue box, is generated in $E_x$ component under *y*-polarized excitation. A reference plane wave will be generated in $E_x$ under *x*-polarized excitation, and its intensity is multiplied by a factor of five for clarity. Similar intensity measurement is taken under 'LCP', 'RCP', '+45°' and '-45°' excitation. FDTD computed and experimentally measured maps show good agreement.

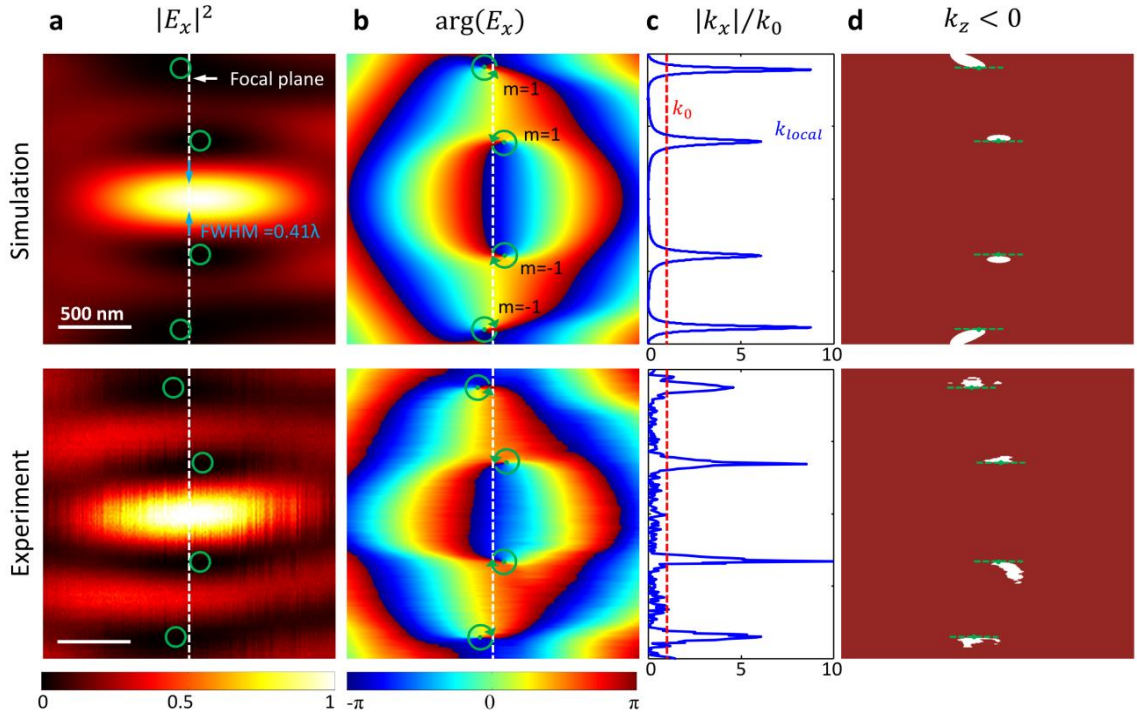

**Figure S6.** Four characteristic features of the superoscillatory field (TM configuration). (Top row) simulation; (Bottom row) experiment. **a**, *High localization of the field* can be seen from the intensity map $|E_x|^2$ near the superoscillatory focus annotated in the blue box in Fig. S4; **b**, *Phase singularities* with topological charge of $m = +1$ and $-1$ are seen on the phase maps $\arg(E_x)$ in the low-intensity areas of the superoscillatory field highlighted by green circles; **c**, *Gigantic local wave-vectors* $|k_x|/k_0$ at the focal plane are calculated from the phase gradient in $x$-direction; **d**, *The energy backflow* areas with negative longitudinal wave-vector $k_z < 0$ are highlighted in white. Dashed green lines indicate tangent to the retro-propagation areas at the point of their intersections with phase singularities.